# Classical simulation of qubit by a new hardware design and its applications

A. Chouikh[1], T. Said[1] and M. Bennai[1,2].

[1]Physics and Quantum Technology Team, LPMC, Faculty of Science Ben M.sik, Casablanca Hassan II University, Morocco

[2]Lab of High Energy Physics, Modeling and Simulations, Faculty of Science, University Mohammed V-Agdal, Rabat, Morocco

abdelhaq.chouikh@gmail.com; taoufik.said81@gmail.com; mdbennai@yahoo.fr

In this paper, we develop a new classical simulation of quantum bit (qubit) by use of analog components in order to be able to simulate the quantum properties such as the superposition of states. As part of this new approach, we have also implemented many logical gates (Pauli-*X*, Hadamard, and Conditional phase-shift gates). Simulation of Hadamard and Conditional phase-shift gates using *Proteus Design Suite* is also presented and discussed in this work.

## I. Introduction

In the classical information theory developed by Shannon [1-2], information [3] is quantified in terms of discrete units of information called bits. Each bit has two possible values 0 and 1. The information lies in specifying which of the two values to assign to the bit. Thus, the information unit bit corresponds to the two-state system as a physical unit. Such a system can be realized in many ways, as a physical system that can easily be switched between two stable states, as an electric signal with only two allowed values "on" or "off". The important point to note is that the two-state system considered in this way is a classical system and the interesting question which has been addressed in recent years is whether quantum physics should introduce a new picture of the (physical) unit of information [4-7]. The classical two-state system has its counterpart in the quantum two-state or two-level system, and for the quantum system a new feature is that coherent superposition between states are possible. In the same way as the classical two-state system is associated with a bit of information, the quantum two-level system is associated with a new information unit, a qubit. While the possible values of a bit is restricted to 0 and 1, the qubit takes values in a two-dimensional Hilbert space spanned by two vectors $|0\rangle$ and $|1\rangle$.

The digital electronics rely on binary logic to store, process, and transmit data or information[8,9]. Binary Logic refers to one of two states – ON or OFF. This is commonly translated as a binary 1 or binary 0. A binary 1 is also referred to as a HIGH signal and a binary 0 is referred to as a LOW signal. The strength of a signal is typically described by its voltage level. How is a logic 0 (LOW) or a logic 1 (HIGH) defined? When below the low threshold, the signal is "low." When above the high threshold, the signal is "high." Intermediate levels are undefined, resulting in highly implementation-specific circuit behavior. Manufacturers of chips generally define these in their spec sheets. The most common standard is TTL [10] or Transistor-Transistor Logic.

It is usual to allow some tolerance in the voltage levels used; for example (TTL Technology), 0 to 0.8 volts might represent logic 0, and 2 to 5 volts logic 1. A voltage of 0.8 to 2 volts would be invalid, and occur only in a fault condition or during a logic level transition.

We notice that for a signal whose potential is between two limit values 0 and Vcc (Vcc = 5v for TTC), there is an infinity of values that can take the electric potential, but we restrict on two intervals only for define two logical levels 0 (low) and 1 (high). I then wondered if we can take advantage of this continuous interval with an infinite number of points to build a logic similar to the quantum logic based on the superposition of two states $|0\rangle$ and $|1\rangle$. Otherwise, instead of restricting a classical state to two voltage intervals (o or 1), we consider each point of the interval as a physical state (bit). That's why we have thought about establishing a new representation "the quantum representation of a classic signal".

It should be noted that some works [23-32] have made use of mechanisms where quantum computation can be achieved by allowing gates or algorithms to be conditioned on classical bits i.e. they are executed on a classical computer. But in this paper, we develop a classical simulation of quantum bit (qubit) by use of analog components; i.e, the hardware used is not binary and the quantum computation will not be realized in a classical computer. This work, to my knowledge, has not been done yet.

## II. Quantum Representation

Quantum states are the key mathematical objects in quantum theory. It is therefore surprising that physicists have been unable to agree on what a quantum state truly represents. One possibility is that a pure quantum state corresponds directly to reality. However, there is a long history of suggestions that a quantum state (even a pure state) represents only knowledge or information about some aspect of reality. Many others have suggested that the quantum state is something less than real [11-19]. In particular, it is often argued that the quantum state does not correspond directly to reality, but represents an experimenter's knowledge or information about some aspect of reality. This view is motivated by, amongst other things, the collapse of the quantum state on measurement. Many will then continue to view the quantum state as representing information.

In this paper, we introduce a new representation of the classical state of an electrical signal. We will quantify classical information in a similar way to that used for a quantum system.

### A. Definition

For any given electrical signal $s_{cl}(t)$ (with domain $\{-1,1\}$), we can define a temporal representation of a fictive quantum state $|\psi(t)\rangle$ such that:

$$|\psi(t)\rangle = s_{qu}(t)|0\rangle + s_{cl}(t)|1\rangle \qquad (1)$$

Where $|\psi(t)\rangle$ is a unit vector in vector space $\mathbb{C}^2$, the coefficients $s_{qu}(t)$ is a complex number and the set $\{|0\rangle, |1\rangle\}$ is the computational basis of $\mathbb{C}^2$. We note that:

- $|1\rangle$ : is the classical state vector of the representative quantum state. It corresponds to the superior limit of voltage.
- $|0\rangle$ : is the quantum state vector (fictive) that corresponds to the central value of voltage.
- $s_{cl}(t)$ : is the classical component of the state $|\psi(t)\rangle$
- $s_{qu}(t)$ : is the quantum component of the state $|\psi(t)\rangle$

### B. Remark 1

For a signal $S_{cl}$ in a given domain $\{f_{min}, f_{max}\}$, we can always reinitialize it in $\{-1,1\}$ using the following transformation:

$$s_{re} \to \frac{2s_{re} - (f_{max} + f_{min})}{(f_{max} - f_{min})} \qquad (2)$$

### C. Remark 2

The knowledge of the function $s_{cl}(t)$, allows us to determine the function $s_{qu}(t)$ by:

$$s_{qu}(t) = \mp e^{-i\phi}\sqrt{1 - s_{cl}^2(t)} \quad (0 \leq \phi < 2\pi) \qquad (3)$$

We have associated our representation to a quantum state of an electrical signal then the wave function has to be normalized since without normalizing, the notion of probability (see conclusion) wouldn't make any sense. Else with assuming that the energy losses are negligible (absence of information loss), any transformation that the signal state undergoes must be unitary then if a wave function is initially normalized it stays normalized. We can therefore apply the normalization condition to the state $|\psi(t)\rangle$:

$$s_{qu}^2(t) + s_{cl}^2(t) = 1 \qquad (4)$$

Which gives $s_{qu}(t)$ according to $s_{cl}(t)$.

### D. Remark 3

The passage from the quantum representation of a signal to its classical representation and vice versa is done by the following correspondence (TABLE1):

TABLE1
CORRESPONDENCE BETWEEN CLASSICAL
AND QUANTUM REPRESENTATION

| Quantum representation | classical signal |
| --- | --- |
| $|1\rangle$ | 1 |
| $|0\rangle$ | 0 |
| $s_{qu}(t)|0\rangle + s_{cl}(t)|1\rangle$ | $s_{cl}(t)$ |

## III. INFORMATION CODING

In the traditional communicational context, whose classical locus is Claude Shannon's formalism [1,2], information is primarily something that has to be transmitted between two points for communication purposes. Shannon's theory is purely quantitative, it ignores any issue related to informational content: Shannon wrote that the "semantic aspects of communication are irrelevant to the engineering problem. The significant aspect is that the actual message is one selected from a set of possible messages"[1].

Otherwise, there are no reasons to admit the existence of quantum information as qualitatively different from classical information: there is only one kind of information, physically neutral, which can be encoded by means of classical or of quantum states [20].

### A. Quantum Quantization of Classical Information

The first papers on quantum models for computation were published in the 1980s. Similar to the research into reversible models, the motivation was mostly academic at the time, an exploration of the ultimate limits of computation. The real payoff for quantum computing did not come until 1994, when Shor announced his quantum algorithm for factoring large numbers with an efficiency unparalleled by any classical algorithm preceding it [21,22]. The factoring problem is used widely to encrypt messages in public key cryptography, which made the feasibility of quantum computing an urgent issue in the years to follow.

The issue we will address here is the following: Is this quantum representation of a classical state able to form a model for computation, and whether there are qualitative differences between such a model and the preceding models for computation?

In this work, although this quantum representation does not give a true picture of the reality, we will use it to give a quantum quantization of classical information. We note the smallest amount of information per cqubit to differentiate it from the other notations (bit and qubit).

$$s_{cl}(t) \dashrightarrow |cqubit\rangle = s_{qu}(t)|0\rangle + s_{cl}(t)|1\rangle \qquad (5)$$

with:  $\qquad s_{qu}(t) = \mp e^{-i\phi}\sqrt{1 - s_{cl}^2(t)} \qquad (6)$

### B. Circuit Representation of Cqubit

In this approach, we restrict ourselves to sinusoidal signals since the signal description in the frequency domain simplifies calculations and also the considered sine wave signals are not subjected a shape variation due to the transformations they undergo. By taking $s_{cl}(t) = \sin(\omega t + \varphi)$, a completely arbitrary, unknown cqubit state is written as:

$$|\psi(t)\rangle = e^{i\alpha}\cos(\omega t + \varphi)|0\rangle + e^{i\beta}\sin(\omega t + \varphi)|1\rangle \qquad (7)$$

Or by taking $\phi = \beta - \alpha$:

$$|\psi(t)\rangle = e^{i\beta}(e^{-i(\beta-\alpha)}\cos(\omega t + \varphi)|0\rangle + \sin(\omega t + \varphi)|1\rangle)$$
$$= e^{-i\phi}\cos(\omega t + \varphi)|0\rangle + \sin(\omega t + \varphi)|1\rangle \quad (8)$$

The factor of $\exp(i\beta)$ is just a global phase, and so was ignored. $\omega$ is the frequency of the electrical signal, $\varphi$ is the initial phase and $\phi$ is the azimuthal angle ($0 \leq \phi < 2\pi$). $|\psi(t)\rangle$ is geometrically represented by the points on the surface of a three dimensional sphere, known as the Bloch sphere [4] and is characterized by the frequency $\omega$ and the azimuthal angle $\phi$. We can represent and generate the cqubit state circuit from a given classical signal $S_{cl}(t) = \sin(\omega t + \varphi)$ by using equation.6. We note that the quantum component $S_{qu}(t)$ can be simulated classically in parallel with the calculation of classical part $S_{cl}(t)$. We can therefore summarize this diagram circuit as follows (see Fig. 1):

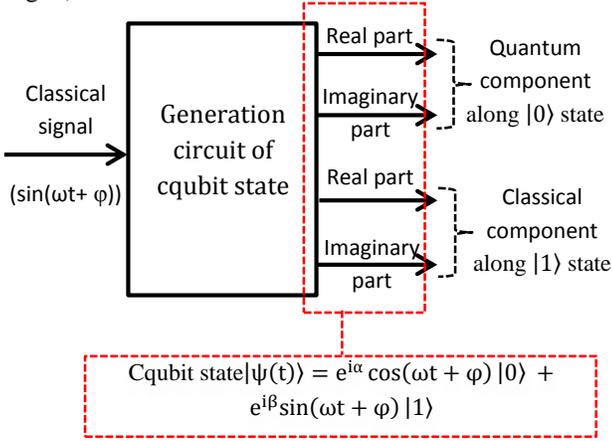

Fig. 1. A schematic diagram circuit of the generation of cqubit state from a given classical signal $\sin(\omega t + \varphi)$ ($|\psi(t)\rangle = e^{i\alpha}\cos(\omega t + \varphi)|0\rangle + e^{i\beta}\sin(\omega t + \varphi)|1\rangle$).

This circuit comprises two parallel blocks. In addition to adding different phases corresponding to the $\alpha$ and $\beta$ angles, the first block leaves the signal $\sin(\omega t + \varphi)$ invariant while the second transforms it into $\cos(\omega t + \varphi)$. This latter can be, for example, a derivative or a ($-\pi/2$) phase-shifter circuit. Four voltage generators can also be used directly, to generating the two sines and the the two cosines.

## IV. LOGIC GATES IMPLEMENTATION

In the following, we will implement some universal logic gates: Hadamard, Pauli-X and Conditional phase-shift gates. For implementing logical gates represented by unitary matrices with complex elements (Conditional Phase-Shift,...) or for any input state ($\phi \neq 0$), we can use two parallel calculating units, which allowed the computer to simultaneously perform calculations on both parts of a complex number [33]. In the main, we will show that any quantum logic gate can be implemented by classical process using cqubit representation. We will describe all the ideas and give several examples.

### A. Hadamard gate implementation

The Hadamard-gate (denoted H) is one of the most useful single-qubit gates, since it turns $|0\rangle$ into $(|0\rangle + |1\rangle)/\sqrt{2}$ and $|1\rangle$ into $(|0\rangle - |1\rangle)/\sqrt{2}$, therefore creating superposition.

$$H = \frac{1}{\sqrt{2}}\begin{pmatrix} 1 & 1 \\ 1 & -1 \end{pmatrix} \quad (9)$$

We Note that H is real and symmetric ($H^2 = I$). In the complex plane, H can be visualized as a reflection around $\pi/8$, or a rotation around $\pi/4$ followed by a reflection. On the Bloch sphere H can also be visualized in several ways. One is a rotation of $\pi/2$ about the y-axis, followed by a rotation about the x-axis by $\pi$. Another is a rotation of $\pi$ about the axis $(1/\sqrt{2}, 0, 1/\sqrt{2})$ [4,34,35]. The operator H acts linearly:

$$H(|\psi(t)\rangle_{in}) = H(e^{i\alpha}\cos(\omega t + \varphi)|0\rangle + e^{i\beta}\sin(\omega t + \varphi)|1\rangle)$$
$$= (1/\sqrt{2})[(e^{i\alpha}\cos(\omega t + \varphi) + e^{i\beta}\sin(\omega t + \varphi))|0\rangle \quad (10)$$
$$+ (e^{i\alpha}\cos(\omega t + \varphi) - e^{i\beta}\sin(\omega t + \varphi))|1\rangle]$$

By writing $|\psi(t)\rangle_{in}$ as:

$$|\psi(t)\rangle_{in} = (C_{in}^{0,Re} + iC_{in}^{0,Im})|0\rangle + (C_{in}^{1,Re} + iC_{in}^{1,Im})|1\rangle \quad (11)$$

with
- $C_{in}^{0,Re}$ and $C_{in}^{0,Im}$ are respectively the real and imaginary part of the quantum component along the $|0\rangle$ state: $C_{in}^{0,Re} = \cos(\alpha)\cos(\omega t + \varphi)$ and $C_{in}^{0,Im} = \sin(\alpha)\cos(\omega t + \varphi)$.
- $C_{in}^{1,Re}$ and $C_{in}^{1,Im}$ are respectively the real and imaginary part of the classical component along the $|1\rangle$ state vector: $C_{in}^{1,Re} = \cos(\beta)\sin(\omega t + \varphi)$ and $C_{in}^{1,Im} = \sin(\beta)\sin(\omega t + \varphi)$.

We found $|\psi(t)\rangle_{out}$ as:

$$|\psi(t)\rangle_{out} = H(|\psi(t)\rangle_{in}) \quad (12)$$
$$= 1/\sqrt{2}\{[\cos(\alpha)\cos(\omega t + \varphi) + \cos(\beta)\sin(\omega t + \varphi)$$
$$+ i(\sin(\alpha)\cos(\omega t + \varphi) + \sin(\beta)\sin(\omega t + \varphi))]|0\rangle$$
$$+ [(\cos(\alpha)\cos(\omega t + \varphi) - \cos(\beta)\sin(\omega t + \varphi))$$
$$+ i(\sin(\alpha)\cos(\omega t + \varphi) - \sin(\beta)\sin(\omega t + \varphi))]|1\rangle\}$$
$$= 1/\sqrt{2}\{[(C_{in}^{0,Re} + C_{in}^{1,Re}) + i(C_{in}^{0,Im} + C_{in}^{1,Im})]|0\rangle$$
$$+ [(C_{in}^{0,Re} - C_{in}^{1,Re}) + i(C_{in}^{0,Im} - C_{in}^{1,Im})]|1\rangle\}$$

Then we can write the components of $|\psi(t)\rangle_{out}$:

$$\begin{cases} C_{out}^{0,Re} = 1/\sqrt{2}(C_{in}^{0,Re} + C_{in}^{1,Re}) \\ C_{out}^{0,Im} = 1/\sqrt{2}(C_{in}^{0,Im} + C_{in}^{1,Im}) \\ C_{out}^{1,Re} = 1/\sqrt{2}(C_{in}^{0,Re} - C_{in}^{1,Re}) \\ C_{out}^{1,Im} = 1/\sqrt{2}(C_{in}^{0,Im} - C_{in}^{1,Im}) \end{cases} \quad (13)$$

The implementation of the Hadamard gate requires therefore two steps: In a first step we implement the expressions $(C_{in}^{0,Re} + C_{in}^{1,Re})$, $(C_{in}^{0,Im} + C_{in}^{1,Im})$, $(C_{in}^{0,Re} - C_{in}^{1,Re})$ and $(C_{in}^{0,Im} - C_{in}^{1,Im})$ in parallel by using four analog electronic components: two adders and two subtractors. In the second step we use four analog voltage dividers to implement the coefficient $1/\sqrt{2}$. We give in Fig. 2, the diagram of Hadamard gate circuit:

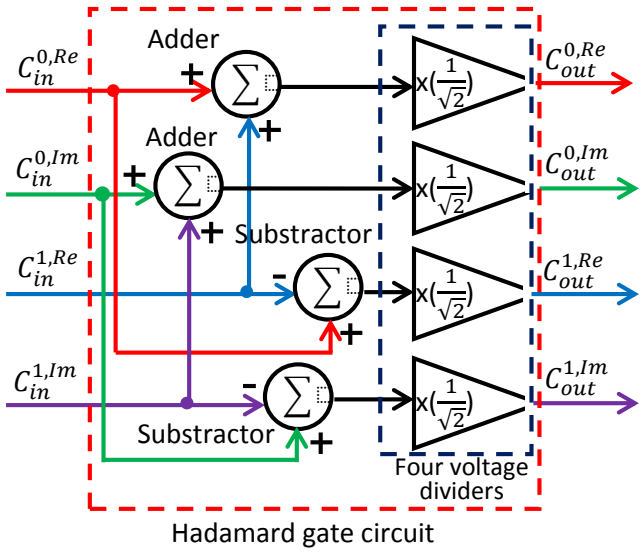

Hadamard gate circuit

Fig. 2; A schematic diagram of implementing *H* gate which operates on the classical and quantum components. An input state $|\psi(t)\rangle$ is prepared and a set of two transformations are applied (Two adders and two subtractors in a first step and four voltage dividers in the second)

## B. Pauli-X gate implementation

The Pauli X-gate corresponds to a classical NOT gate. For this reason, in quantum information the X-gate is often called the quantum NOT gate as well. Here's what X matrix looks like:

$$X = \begin{pmatrix} 0 & 1 \\ 1 & 0 \end{pmatrix}$$

The operator X acts linearly:

$$H(|\psi(t)\rangle_{in}) = H(e^{i\alpha}\cos(\omega t + \varphi)|0\rangle + e^{i\beta}\sin(\omega t + \varphi)|1\rangle)$$
$$= e^{i\beta}\sin(\omega t + \varphi)|0\rangle + e^{i\alpha}\cos(\omega t + \varphi)|1\rangle \quad (14)$$

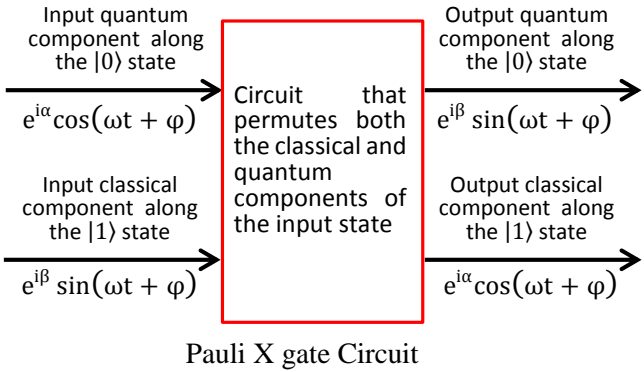

Pauli X gate Circuit

Fig. 3. A schematic diagram of implementing Pauli **X** gate which operates on the classical and quantum components. An input state $|\psi(t)\rangle_{in}$ is prepared and a transformation that permutes both the classical and quantum components of the input state are applied.

So the problem is finding a circuit that turns $e^{i\alpha}.\cos(\omega t+\varphi)$ to $e^{i\beta}.\sin(\omega t+\varphi)$ and $e^{i\beta}.\sin(\omega t+\varphi)$ to $e^{i\alpha}.\cos(\omega t+\varphi)$. This transformation can be implemented by various analog circuits, for example by permutation of the two components (classical and quantum).

We give in Fig. 3, the circuit diagram of the X gate circuit.

## C. Conditional Phase-Shift gate implementation

The conditional phase-shift gate introduces a phase shift only if a predetermined condition is satisfied. In one-qubit system, the conditional phase-shift gate is an unitary transform of the form:

$$R_\phi = \begin{pmatrix} 1 & 0 \\ 0 & e^{i\phi} \end{pmatrix}$$

The $R_\phi$ introduces a phase-shift if the qubit is in state $|1\rangle$. The operator $R_\phi$ acts linearly as:

$$R_\phi(|\psi(t)\rangle_{in}) = R_\phi(e^{i\alpha}\cos(\omega t + \varphi)|0\rangle + e^{i\beta}\sin(\omega t + \varphi)|1\rangle)$$
$$= e^{i\alpha}\cos(\omega t + \varphi)|0\rangle + e^{i\phi}e^{i\beta}\sin(\omega t + \quad (15)$$

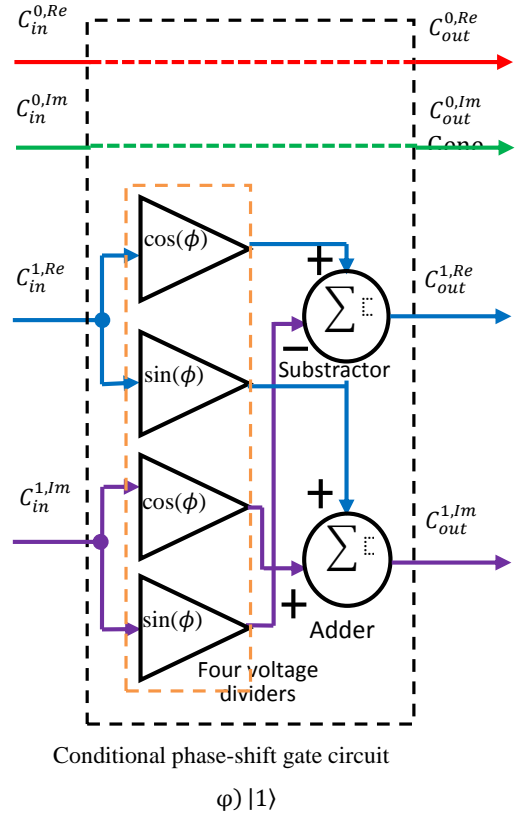

Conditional phase-shift gate circuit

$\varphi)|1\rangle$

Fig.4. A schematic diagram of implementing Conditional phase shift gate. An input state $|\psi(t)\rangle$ is prepared and a set of two transformations are applied (Four analog voltage divider circuits to implement the coefficients $\cos(\phi)$ and $\sin(\phi)$ in a first step and in a second step one analog adder and one analog subtractor circuits to implement respectively the sum and the subtraction in equation 18).

The quantum component along the $|0\rangle$ state remains unchanged and by writing $|\psi(t)\rangle_{in}$ as:

$$|\psi(t)\rangle_{in} = (C_{in}^{0,Re} + iC_{in}^{0,Im})|0\rangle + (C_{in}^{1,Re} + iC_{in}^{1,Im})|1\rangle \quad (16)$$

we found the classical component along the state vector $|1\rangle$:

$$e^{i\phi}e^{i\beta}\sin(\omega t + \varphi) = [\cos\phi\cos\beta\sin(\omega t + \varphi) - \sin\phi\sin\beta\sin(\omega t + \varphi)]$$
$$+ i[\cos\phi\sin\beta\sin(\omega t + \varphi) + \sin\phi\cos\beta\sin(\omega t + \varphi)]$$
$$= [\cos(\phi)C_{in}^{1,Re} - \sin(\phi)C_{in}^{1,Im}] + i[\cos(\phi)C_{in}^{1,Im} + \sin(\phi)C_{in}^{1,Re}] \quad (17)$$

Finally, we can write the components of $|\psi(t)\rangle_{out}$ as:

$$\begin{cases} C_{out}^{0,Re} = C_{in}^{0,Re} \\ C_{out}^{0,Im} = C_{in}^{0,Im} \\ C_{out}^{1,Re} = \cos(\phi)C_{in}^{1,Re} - \sin(\phi)C_{in}^{1,Im} \\ C_{out}^{1,Im} = \cos(\phi)C_{in}^{1,Im} + \sin(\phi)C_{in}^{1,Re} \end{cases} \quad (18)$$

Fig.4 give us, the circuit diagram of the conditional phase-shift gate:

## V. SIMULATION OF HADAMARD AND CONDITIONAL PHASE-SHIFT GATES USING PROTEUS DESIGN

Proteus software is developed by Labcenter Electronics. Proteus is software used for electronic circuits, microprocessor based circuits simulation and for designing printed circuit board (PCB). The main feature of Proteus design software is its multiple system components. It is used to draw schematics and the simulation allows human access during run time, thus providing real time simulation. In this paragraph we will test our model taking Hadamard and conditional phase-shift gates as an example. We will use proteus to perform this simulation.

### A. Hadamard Gate

As already seen, implementation of Hadamard gate requires two steps: in first step we use the adders (S2, S5) and the subtractors (S1, S6) circuits and in second step, four voltage dividers allows us to implement the coefficient $1/\sqrt{2}$. To provide a stiff signal we use a follower (operational amplifier) whose base is powered by the voltage divider.

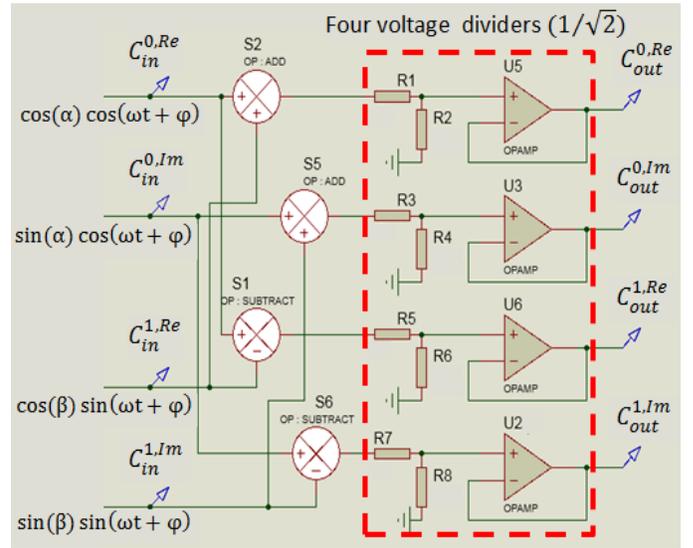

Fig. 5. Simulation circuit of Hadamard gate

The implementation circuit simulated using Proteus is shown Fig. 5. The design parameters and the components values are presented in TABLE2. The operating frequency of the signal is considered as 1 GHz which corresponds to an angular frequency ω of the order of $2\pi.10^9$ rad.s$^{-1}$. We note that we can set the input state by changing the value of φ, α and β.

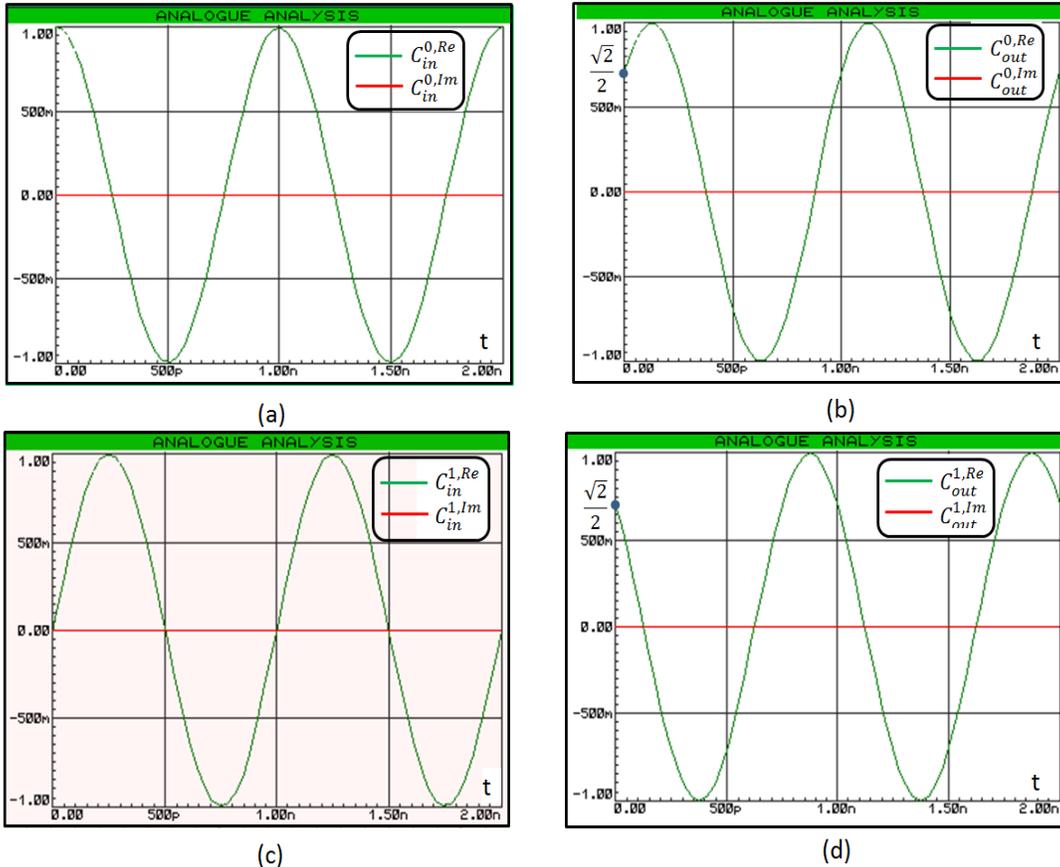

Fig. 6. Waveform of input and output voltages vs time (in ns) for Hadamard gate circuit: (a) and (c) present respectively the quantum and classical input components along the |0⟩ and |1⟩state vectors while (b) and (d) present respectively the quantum and classical output components along the |0⟩ and |1⟩ state vectors. Green and red colors mean respectively real and imaginary parts.

Here, we analyse the Hadamard circuit. For that, we take as an example φ=0, α=0 and β = 0. From these design values we have $C_{in}^{0,Re}(t) = \cos(\omega t)$, $C_{in}^{1,Re}(t) = \sin(\omega t)$, $C_{in}^{0,Im}(t) = 0$ and $C_{in}^{1,Im}(t) = 0$. We give in Fig. 6.a and Fig. 6.c the curves associated with input voltages $C_{in}^{0,Re}(t)$, $C_{in}^{1,Re}(t)$, $C_{in}^{0,Im}(t)$ and $C_{in}^{1,Im}(t)$. Fig. 6.b and Fig. 6.d shows the waveform for output voltages $C_{out}^{0,Re}(t)$, $C_{out}^{1,Re}(t)$, $C_{out}^{0,Im}(t)$ and $C_{out}^{1,Im}(t)$.

TABLE2
VALUES OF THE CIRCUIT PARAMETERS

| Parameters | Values (kΩ) | Parameters | Values (kΩ) |
|---|---|---|---|
| Resistor (R1) | 10 | Resistor (R2) | 24.14 |
| Resistor (R3) | 10 | Resistor (R4) | 24.14 |
| Resistor (R5) | 10 | Resistor (R6) | 24.14 |
| Resistor (R7) | 10 | Resistor (R8) | 24.14 |

We see that in Fig. 6.b, $C_{out}^{0,Re}(t)$ corresponds to $(1/\sqrt{2})(\cos(\omega t) + \sin(\omega t))$ and $C_{out}^{0,Re}(t) = 0$. While in Fig. 6.d, $C_{out}^{1,Re}(t)$ corresponds to $(1/\sqrt{2})(\cos(\omega t) - \sin(\omega t))$ and $C_{out}^{1,Im}(t) = 0$. These relationships coincide exactly with the theoretical values of equation 13. We have do tests on other values of input state to check the performance of the Hadamard gate circuit.

*B. Conditional Phase-Shift Gate*

In the same way we implement the conditional phase-shift gate. We note that quantum components along the state vector |0⟩ remains unchanged while classical components along the state vector |1⟩ requires two steps: in a first step we use four voltage divider to implement the coefficients cos(ϕ) and sin(ϕ) and in a second step one adder (S2) and one subtractor (S1) circuits allows us to implement respectively the expressions $\cos(\phi). C_{in}^{1,Im} + \sin(\phi). C_{in}^{1,Re}$ and $\cos(\phi). C_{in}^{1,Re} - \sin(\phi). C_{in}^{1,Im}$. The operating frequency of signal is considered as 1 GHz (ω=2π.10⁹ rad.s⁻¹). We set the input state by changing the value of φ, α and β. For the value of ϕ we take the case of the π/8 gate where ϕ = π/4. Then cos(ϕ) = sin(ϕ) = 1/√2. We use the same components values of TABLE2 to realize four divider circuits having a 1/√2 coefficient.

The implementation circuit simulated using Proteus is shown in Fig. 7

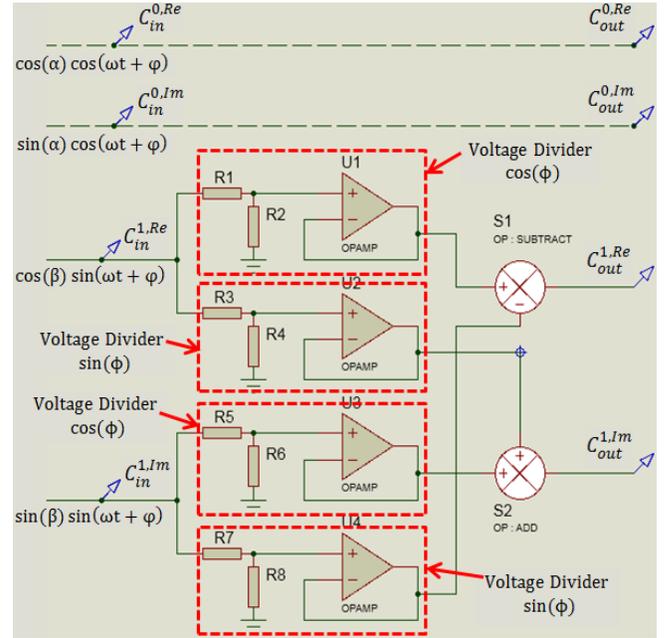

Fig. 7. Simulation circuit of conditional phase-shift gate

We analyse only classical components along the state vector |1⟩ since quantum components along the state vector |0⟩ remains unchanged. For it we take as an example φ = π/2 and β = π/2. From these design values we have $C_{in}^{1,Re}(t) = 0$ and $C_{in}^{1,Im}(t) = \sin(\omega t + \pi/2)$. We give in Fig.8.a the curves associated with input voltages $C_{in}^{1,Re}(t)$ and $C_{in}^{1,Im}(t)$. Fig.8.b show the waveform for output voltages $C_{out}^{1,Re}(t)$ and $C_{out}^{1,Im}(t)$. We see that in Fig.8.b, $C_{out}^{1,Re}(t)$ correspond to $(-1/\sqrt{2}) \sin(\pi/2) \sin(\omega t + \pi/2)$ and $C_{out}^{1,Im}(t)$ correspond to $(1/\sqrt{2}) \sin(\pi/2) \sin(\omega t + \pi/2)$. These relations coincide exactly with the theoretical values in equation 18. We have do tests on other values of input state to check the performance of the conditional phase-shift gate circuit.

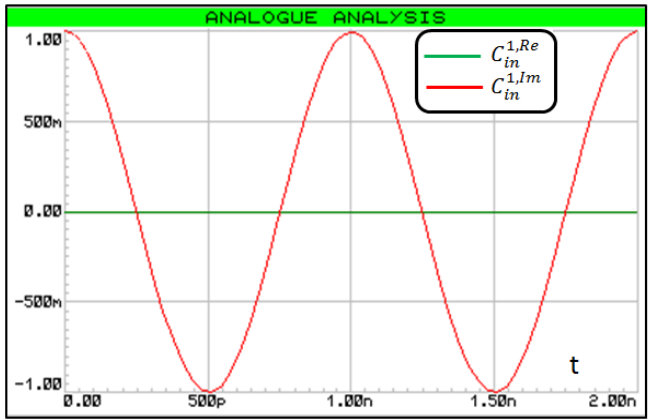 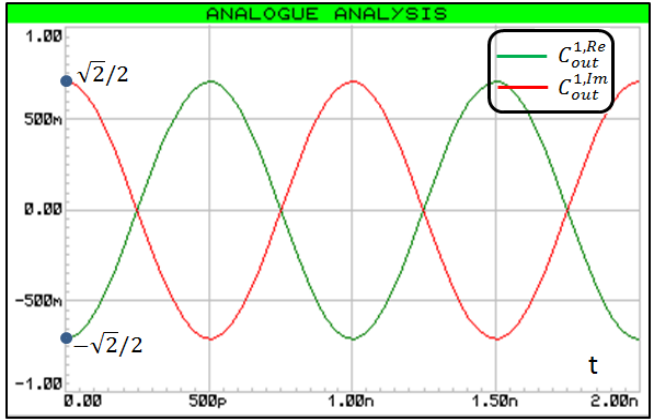

Fig.8. Waveform of input and output voltages vs time (in ns) for conditional phase-shift gate circuit: (a) present the input classical components along $|1\rangle$ state vector while (b) present the output classical components along the $|1\rangle$ state vector. Green and red colors mean respectively real and imaginary parts

## VI. CONCLUSION

In conclusion, we have described a new scheme of computation in which we have taken advantage of a signal voltage in the continuous interval [-$V_{cc}$, $V_{cc}$] with an infinite number of points to build a logic similar to the quantum logic which is based on the superposition of two states.

For this, we have established a new representation that we have named the quantum representation of a classical signal. The ability to implement quantum computation with classical processes may allow to realize the dream of a "quantum" computer, freed from the constraints of classical computers' 0s or 1s, using a classic physical process that can deliver optimal results faster or can too overcome the different engineering challenges [36-41] when building a quantum computer ranging from the core qubit technology, the control electronics, to the microarchitecture for the execution of quantum circuits, efficient quantum error correction and the great problem of decoherence.

Simulation of Hadamard and Conditional phase-shift gates using Proteus Design Suite is also presented and discussed in this paper. The circuits are designed with the real time hardware components and the results are generated for sinusoïdal input voltages. The curves of output voltages are obtained for specific input voltages and the analysis of these curves shows that the implementations of these two circuits correspond to Hadamard and Conditional phase-shift gates. Finally, we note some remarks:

- $S_{cl}$ and $S_{qu}$ represent voltages, but since the maximum value is the unit, we can consider them without dimension (divide it by the maximum value) and their squares represent a probability. It expresses the percentage probability that the voltage limit has been reached.
- In quantum information, we take into account the implementation time of a circuit which must be shorter than decoherence time and relaxation time of a qubit. While classical circuits use a time named propagation delay (symbolized $t_{pd}$) which is the time required for a digital signal to travel from the input of a logic gate to the output. In our scheme the time required for a signal to moving through a gate is almost instantaneous but must be considered for complex circuits.

**Author Contributions**

A.C. proposed the idea, developed the theory and the calculations., and was mainly responsible for writing the manuscript. T.S. contributed to the conception of electronic circuits and results analysis. M.B. provided feedback and supervised the project. All authors discussed and reviewed the final version of the manuscript.

**Competing Interests**

The authors declare that they have no competing interests.

**Data availability**

The data that supports the findings of this study are available within the article.